\documentclass[superscriptaddress,twocolumn,aps,prb,floatfix,showpacs]{revtex4}
\usepackage{epsfig}
\usepackage{dcolumn}
\usepackage{bm}
\begin{document}

\title{On the large role of weak uncorrelated pinning introduced by BZO
nanorods at low temperatures in REBCO thin films}

\author{A. Xu}
\email{aixiaxu@magnet.fsu.edu}\affiliation{National High Magnetic
Field Laboratory, Florida State University, Tallahassee, Florida
32310}
\author{V. Braccini}
\affiliation{CNR-SPIN, Corso Perrone 24, I-16152 Genova, Italy}
\author{J. Jaroszynski}
\affiliation{National High Magnetic Field Laboratory, Florida
State University, Tallahassee, Florida 32310}
\author{Y. Xin}
\affiliation{National High Magnetic Field Laboratory, Florida
State University, Tallahassee, Florida 32310}
\author{D. C. Larbalestier}
\affiliation{National High Magnetic Field Laboratory, Florida
State University, Tallahassee, Florida 32310}

\date{\today}

\begin{abstract}
REBa$_{2}$Cu$_{3}$O$_{x}$ films can achieve remarkably high
critical current density values by the incorporation of insulating
nanoparticles. A particular interesting case concerns BaZrO$_{3}$
(BZO) nanorods, whose strongly correlated effect is seen at high
temperatures. Here we investigate the field, temperature and
angular dependence of the critical current density over a wide
temperature range from 4.2 K to 77 K, and magnetic fields up to 31
T. We show that the correlated $c$-axis pinning of BZO nanorods
becomes progressively less obvious at lower temperature. Indeed at
4.2 K and fields up to 31 T, the only correlated pinning is for
fields parallel to the film plane. We interpret the change as
being due to significant contributions from dense but weak pins
that thermal fluctuations render ineffective at high temperatures
but which become strong at lower temperatures.
\end{abstract}


\maketitle

\section{INTRODUCTION}

REBa$_{2}$Cu$_{3}$O$_{x}$ (REBCO, where RE = rare earth) thin
films are very promising high temperature superconductors for
applications because of their high current-carrying capability in
very strong magnetic fields over a wide operating temperature
regime. Higher and less anisotropic critical currents $I_{c}$
still remain very desirable for widespread applications.
\cite{David01} Fortunately, it has been demonstrated that several
approaches work for the $I_{c}$ improvement, including increasing
the thickness of the REBCO layer, \cite{Judith09,Glyantsevl11}
mitigating the weak-link effect of grain boundaries
\cite{Daniels00,Hammer01,Feldmann07} and by better understanding
and more effective enhancement of the pinning mechanisms.
\cite{Blatter94}

Enhancement of flux pinning by optimization of nanoscale defects
is fundamental to $J_{c}$ improvement.
\cite{David01,Blatter94,Foltyn07, Holesinger01} More specifically,
it was first established that oxygen deficiencies were the
governing pinning centers in high quality YBCO single crystals.
\cite{Daeumling90,Vargas01} In turn, dislocations and point
defects were found to be strong pins at low temperatures and
important for the typically observed one or two order of magnitude
higher $J_{c}$ of YBCO thin films. \cite{Dam99, Hylton90} Compared
to the growth defects mentioned above, intentionally introduced
pinning centers have captured more attention due to their high
tunability. In early study of the pinning effects of heavy-ion
irradiation on YBCO single crystals large $J_{c}$ increases at all
temperature and magnetic fields, especially for magnetic field
parallel to the irradiation direction was found. \cite{Civale91}
Later, self-assembled BaZrO$_{3}$ nanorods lying along the
$c$-axis were incorporated into YBCO thin films and coated
conductors. BZO has a high lattice mismatch, a high melting
temperature, and it is insoluble in REBCO, all of which contribute
to the substantial improvement of $J_{c}$ over a wide angular
range far from the $c$-axis. \cite{Judith04} A series of Ba-metal
oxide (BMO, where M = Sn, Hf and Ir) second phases acting as
correlated $c$-axis pinning effects were explored thereafter in
order to decrease the anisotropic $J_{c}$.
\cite{Yasunaga08,Hanisch05,Engel07} Unfortunately, the additions
of BMO are detrimental to $T_{c}$ and thus degrade the properties
at liquid nitrogen temperature. However, Harrington {\em et al.}
\cite{Harrington09}  obtained excellent pinning properties while
keeping $T_{c}$ at 92 K even with very high concentrations of
self-assembled RE$_{3}$TaO$_{7}$ (RE = Er, Gd and Yb) nanorods,
where, unlike the tensile stress-inducing BZO, the
RE$_{3}$TaO$_{7}$ introduces a compressive stress on the REBCO.
\cite{Harrington09} More recently, an 0.5 $\mu $m thick pulsed
laser deposition (PLD) YBCO film with embedded double perovskite
Ba$_{2}$YNbO$_{6}$ nanorods showed a maximum flux pinning force
density $F_{p\thinspace max}$ in excess of 30 and 120 GN/m$^{3}$
at 75.5 K and 65 K. \cite{Feldmann10} Such $F_{p\thinspace max}$
values exceed a long challenged benchmark, that of 18-20
GN/m$^{3}$ found in the best pinning-tailored NbTi wire at $T =
4.2$~K. \cite{Meingast89}

In general the maximum $J_{c}$ of any superconductor is limited by
the depairing current density $J_{d} =
{\phi}_{0}/3\sqrt{3}\pi\mu_{0} \lambda^{2}\xi $ according to
anisotropic Ginzburg-Landau theory. \cite{Blatter94, Gurevich07}
Here $\phi_{0}$ is the flux quantum, $\lambda $ is the London
penetration length, $\xi $  is the coherence length and $\mu_{0}$
is the vacuum permeability. In fact REBCO films can clearly exert
strong pinning, as shown by estimates that $\sim $30{\%} of
$J_{d}$ is achieved in YBCO thin films, at least at low fields.
\cite{Miura11} But high field magnets do require further $J_{c}$
improvement by flux pinning engineering of REBCO  conductors. In
spite of REBCO having by far the highest irreversibility fields
$H_{irr}(T)$  of all cuprate superconductors, neither $H_{irr}(T)$
nor $J_{c}(H)$ are high enough for making magnets of $5-10$~T
unless the temperature is reduced below $\sim 30-40$~K. So far,
most pinning studies have been concentrated on $J_{c}$ enhancement
at high temperatures, 65 K to 77 K, and magnetic fields near
self-field for power transmission. However, potential applications
of REBCO conductors extend over much broader temperature and
magnetic field regimes. \cite{David01,Weijers10} Recently, thanks
to the availability of REBCO  conductor with suitable mechanical
properties, the development of all-superconducting magnets above
25~T at 4.2 K became feasible, and some prototypes greater than
30~T were demonstrated. \cite{Weijers10,Markiewicz10,Trociewitz11} However,
coil design depends on understanding the detailed angular
dependence of $J_{c}$ over broad temperature and field ranges so
that safe coil quench design can be predicted with confidence.
Moreover, superconducting magnetic storage systems, motors and
generators, working at intermediate temperatures $\sim $30 K, are
important potential applications too.
\cite{Li11,Nagao08,Iwakuma08} Thus, a systematic study of the
pinning mechanisms over a broad range covering all applications
regimes is indispensible for both practical and fundamental
reasons.

BZO nanorods are well known to produce strong correlated $c$-axis
pinning at high temperatures. Surprisingly, however, recent studies
of samples containing BZO nanorods showed no signs of $c$-axis
correlated pinning at 4.2 K, even though $J_{c}(H)$ was strongly
enhanced. \cite{Xu10,Braccini11} Here we show that the dominant
4.2 K pinning characteristic, valid up to at least 31 T, can be
fit to a standard anisotropic mass scaling except near the
$ab$-plane where correlated pinning, probably by the CuO
charge-reservoir layers enhances $J_{c}$. By measuring the
angular-dependent current density $J_{c}$({$\theta $}) over a wide
range of field and temperature, we observe that this uncorrelated
pinning is valid only below $\sim $30 K. We conclude that these
low temperature angular-independent pins are largely point pins
induced by the strain fields of the BZO nanorods. Because they are
point pins,  they are easily thermally depinned at higher
temperatures, leaving the strong correlated $c$-axis pinning
effects of the BaZrO$_{3\thinspace }$nanorods then quite evident.
At 4.2 K, however, the point pins contribute almost half of the
$J_{c}$

\section{EXPERIMENTAL DETAILS}

We performed an extensive angular $J_{c}$({$\theta $}, $T$, $H)$
characterization of a recent REBCO
 thin film in fields up to 31 T and temperatures
from 4.2 to 77 K. The 1.1~$\mu$m thick film was grown by
metal-organic chemical vapor deposition (MOCVD) on a high strength
metal alloy tape commercially available as Hastelloy, on which a
buffer layer textured by ion-beam assisted deposition (IBAD) was
deposited. A $\sim $2 $\mu $m thick sputtered silver layer was
deposited on the REBCO layer as protection and intermediate
electrical contact layer for the $\sim $50 $\mu $m thick copper
layer, which was electro-plated on it. \cite{Selva11} This sample
is representative of the most advanced coated conductor made by
SuperPower Inc. The Zr addition produces BZO nanorods with an
equivalent flux density $B_{\phi }\approx 3$~T. Earlier study has
shown that such films have high $J_{c}$ properties at both high
and low temperatures. \cite{Braccini11}

The 4.2 K and high field four-probe critical current measurements
were performed in a 52~mm cold bore 15~T superconducting magnet
and the 52~mm warm bore 31~T Bitter magnet, fitted with a 38 mm
bore liquid He cryostat. The 10 K to 77 K measurements were
carried out in a 16 T Physical Property Measurement System (PPMS).
Samples were rotated with respect to the external magnetic field
around the axis parallel to the current direction  to maintain a
maximum Lorentz force configuration. The angle {$\theta $} = 0 is
defined as the applied magnetic field perpendicular to the tape
plane which is parallel to the crystallographic $c$-axis direction
with a typical uncertainty of $1-4^{\circ}$ caused by an offset
caused by the IBAD process. \cite{Xu10, Braccini11, Selva11}

Due to the high critical current ($I_{c})$ values observed at
lower temperatures, samples with different geometries were
prepared in order to avoid harmful Joule heating and overstressing
by the large Lorentz forces ($I_{c}\times B$) possible in
different regimes of temperature and magnetic field. A 50 $\times
$ 500 $\mu $m bar was cut by Nd-YAG (yttrium aluminum garnet)
laser for the $J_{c}$ measurement at 77 K. Even narrower samples,
$\sim $ 10 $\mu $m wide and 200 $\mu $m long were patterned by
SEM/FIB so as to restrict $I_{c}$ to $\le $ 5 A when the sample
was measured in helium gas between 10 and 70~K. In both cases,
copper and silver layers were removed by wet-etching. Larger
bridges about $\sim $ 1 mm wide and 1 cm long were patterned
leaving the silver and copper layers present for the 4.2 K
measurements. Two different home-made $I_{c}$ probes equipped with
rotating sample platforms were used. One had a maximum
current-carrying capability of $\sim 500$~A for high $I_{c}$
measurement in liquid helium, while the second had $\sim 5$~A
capability in the PPMS cryostat for studies at temperatures above
10 K.

TEM images were taken in a JEOL JEM 2011 transmission electron
microscope. The critical temperature ($T_{c})$ is defined as the
temperature where resistance $R$ equals zero. The 77 K
irreversibility field was determined from the field dependence of
$J_{c}$ with the criterion $J_{c}$ = 100 A/cm$^{2}$ For lower
temperatures where $H_{irr}(T)$ is greater than magnetic field
available $H_{irr}(T)$ was assessed from the formula in
\cite{Chen09}.

\section{RESULTS}

The MOCVD sample under study has critical temperature
$T_{c}=90.7$~K.  The nominal composition of this sample is
Y$_{0.6}$Gd$_{0.6}$Ba$_{2}$Cu$_{2.3}$O$_{x}$ with 7.5 at.\ {\%} Zr
doping, composition found to give the highest in-field $J_{c}$
values at 77 K. \cite{Selva11} The sample has $J_{c}$ as high as
3.4 MA/cm$^{2}$ at self-field and 1.0 MA/cm$^{2\thinspace }$at 1
T, at $\theta  = 0^{\circ}$. Even more importantly, both the
irreversibility field, $H_{irr} = 10.2$~T, and the maximum flux
pinning force, $F_{p\thinspace max}$ =12 GN/m$^{3}$ along the
$c$-axis are substantially higher than that of other REBCO films
with similar thickness, $\sim $ 1 $\mu $m.
\begin{figure}[htb]
\includegraphics[width=8.1cm]{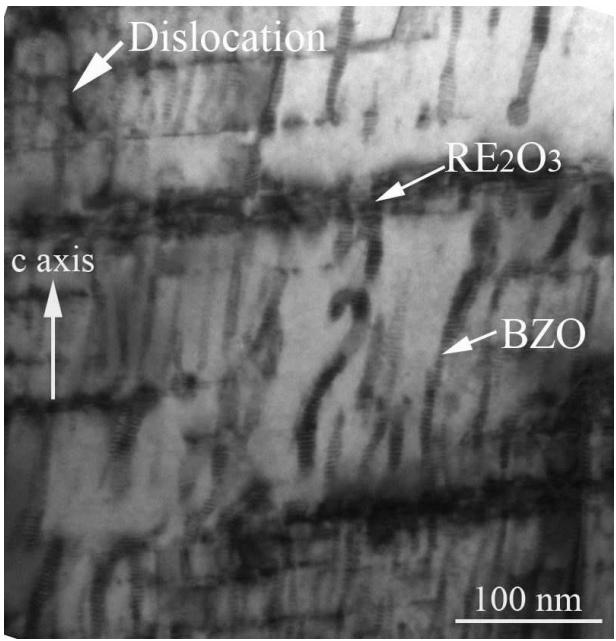}
\caption {Cross-sectional TEM image of the MOCVD sample. Splayed
BZO nanorods along $c$-axis and tilted RE$_{2}$O$_{3}$ precipitate
arrays along $ab$-plane are the major visible correlated pinning
centers. The RE$_{2}$O$_{3}$ precipitates are effective 3D pins
that enhance $J_{c}$ at all orientations. A low density of
threading dislocations that are effective pins along the $c$-axis
are also visible in this image.}\label{fig:RvT}
\end{figure}

Figure 1 shows a typical cross-sectional, bright-field TEM image
of this MOCVD sample. BZO nanorods, with $\sim $8 nm diameter,
with inclinations of $\sim 5-20^{\circ}$ to the $c$-axis are quite
obvious. Their density corresponds to a matching field $B_{\phi }
= \phi_{0} /a^{2}\sim 2.6$~T, where $a
 \approx  28$ nm is the measured average distance between the BZO nanorods
and $\phi_{0}=h/2e \approx  2.1\times 10^{-15}$ Wb is the flux
quantum. Such nanorods are responsible for the broad $I_c$ maxima
observed at elevated temperatures when $H$ is off the film plane.
This strong  correlated pinning  produces the outstanding
superconducting performance observed at 77 K. Moreover, a high
density of self-assembled RE$_{2}$O$_{3}$ precipitate arrays
aligned along the {$ab$}-plane is another important source of
pinning. \cite{Chen09, chen07, Song06} Interestingly, TEM
observation indicates that the {$ab$}-plane of REBCO tilts $\sim
2^{\circ}$ from the buffer layer because of the IBAD process while
the RE$_{2}$O$_{3}$ precipitate arrays tilt away from the
{$ab$}-plane by $\sim 5^{\circ}$, as previously reported.
\cite{Chen09} Threading dislocations provide additional $c$-axis
correlated pinning, although we believe that, their contribution
is negligible compared to BZO nanorods because of their low
density.

Figure 2 (a) presents the $c$-axis field dependence $J_{c}(H||c)$
for magnetic fields up to 16 T or 31 T at various temperatures
from 4.2 K to 77 K. For $T = 77$~K, only data below the
irreversibility field, $H_{irr} = 10.2$~T are plotted. $J_{c}(H)$
shows less field dependence with decreasing temperature. At 10 K,
$J_{c}$ at self-field reaches 43 MA/cm$^{2}$ which corresponds to
$\sim 17$~{\%} of $J_{d}$. $J_{c}$ at 16 T and 10 K is as high as
3.7 MA/cm$^{2}$ equal to the self-field $J_{c}$ at 77 K. At 4.2 K,
$J_{c}$ decreases from 33.3 MA/cm$^{2}$ at 1 T to 2.9 MA/cm$^{2}$
at 31 T. These high current densities correspond to $I_{c}$ = 1.5
kA at 1 T and 0.13~kA at 31 T for standard producion 4 mm wide
tape. Such high $I_c$ values make effective characterization of
such conductors difficult. Notably, this MOCVD BZO-containing
sample shows higher $J_{c}$ at all temperatures below $\sim $70 K
than optimized NbTi wire evaluated at 4.2 K. \cite{Meingast89} The
power-law dependence of $J_{c}$ on magnetic field, $J_{c} \propto
H^{-\alpha }$, is observed at low temperatures.  At 4.2 K, the
power-law exponent $\alpha  = 0.7$ describes $J_{c}$ well in the
whole $1-31$~T range of magnetic field.

\begin{figure}[htb]
\includegraphics[width=8.1cm]{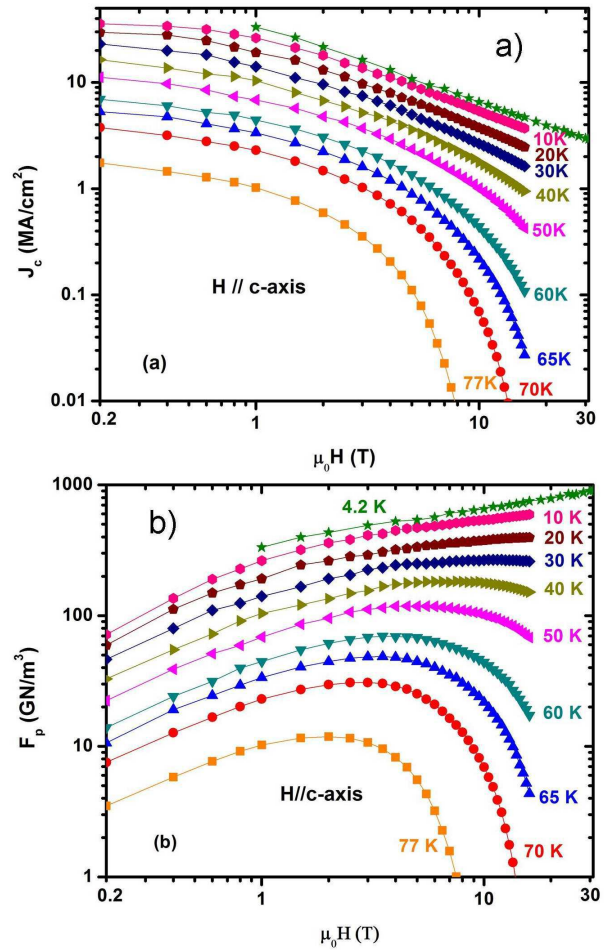}
\caption {(color online)  (a) Field dependence of $J_{c}$ for $H$
parallel to the $c$-axis at temperatures between 4.2 K and 77 K
and magnetic fields up to 31 T and (b) The corresponding flux
pinning force calculated from $F_{p}=J_{c} \times  \mu_0 H$. Only
data below $H_{irr}$ is shown at $T = 77$~K in (a). $J_{c} \propto
H^{-\alpha }$ is observed at temperatures below $\sim 30$~K. It
can be seen that $F_{p\thinspace max}$ is trending to more than 1
TN/m$^{3}$ at the lowest temperature.  Lines connecting data
points are guides for the eyes.}\label{fig:fig2}
\end{figure}

Figure 2 (b) shows flux pinning force density $F_{p}=J_{c} \times
\mu_0 H$ as a function of external magnetic field parallel to the
$c$-axis at various temperatures. The superior $F_{p}$ values
confirm the strong pinning provided by the defects existing in
this sample. The highest measured $F_{p}$ at 4.2 K and 31 T almost
reaches 1000 GN/m$^{3}$. This is the record high value observed in
any superconductor so far. It is also striking that the pinning
force above 4 T at 30 K barely depends on magnetic field. The
maximum pinning force is $\sim $267 GN/m$^{3}$ at 30 K while at
lower temperatures $F_{p}$ maxima correspond to fields higher than
16 T (or even 31 T at 4.2 K) and could not be observed
experimentally. The double-peaked $F_{p}(H)$ dependence shown by
BZO-doped YBCO PLD thin films on SrTiO$_{3}$ single crystal
substrates \cite{Augieri10} are not observed in the present work
at any temperature.

The angular dependence of $J_{c}$ is a powerful tool for
understanding pinning mechanisms and also crucial for magnet
design. Figure 3 presents $J_{c}$({$\theta $, }1 and 4 T) of the
MOCVD sample at 77 K, 50 K, 30 K and 10 K. $J_{c}$ maxima around
the $c$-axis are clearly seen at 77 K and 50 K. These maxima do
not occur exactly at $0^{\circ}$ due to the splayed inclination of
the BZO nanorods. At 1 T and 77 K, the $c$-axis $J_{c}$ reaches
1.1 MA/cm$^{2}$, about one third higher than the $J_{c}$ maximum
value around the {$ab$}-plane and twice the minimum $J_{c}$ close
to the {$ab$}-plane. As the magnetic field increases to 4 T, the
$c$-axis $J_{c}$ peak becomes lower than the {$ab$}-peak, because
the field exceeds the matching field of 2.6 T, corresponding to
the BZO nanorod density. It is worth noting that both BZO nanorods
and the RE$_{2}$O$_{3}$ precipitate arrays contribute to $J_{c}$
at 77 K, being responsible for the $c$-axis peak and for $J_{c}$
enhancement over the whole angular range especially around the
{$ab$}-plane. Comparing the 77 K to lower temperature data at the
same field, the $c$-axis peak becomes less distinct as temperature
decreases, not being observable at all at temperatures below 30 K.
This strongly suggests that the dominant pinning mechanism changes
at lower temperatures and that the crossover temperature is $\sim
$30 K. The ratio of $J_{c}||c$ and $J_{c}||{ab}$ decreases from
1.4 at 77 K to 0.9 at 50 K, showing the reversion to that expected
by the mass anisotropy at low temperature. At lower temperature,
the $J_{c}||{ab}$ peak becomes more evident, indicating a
strengthening of the {$ab$}-plane correlated pinning. At low
fields $J_{c}||{ab}$ varies from a cusplike dependence above 50 K
to a smooth, Ginzburg-Landau (GL)-like peak at lower temperatures
and at 10 K from GL-like at low fields to cusp-like at high
magnetic fields.

\begin{figure}[htb]
\includegraphics[width=8.1cm]{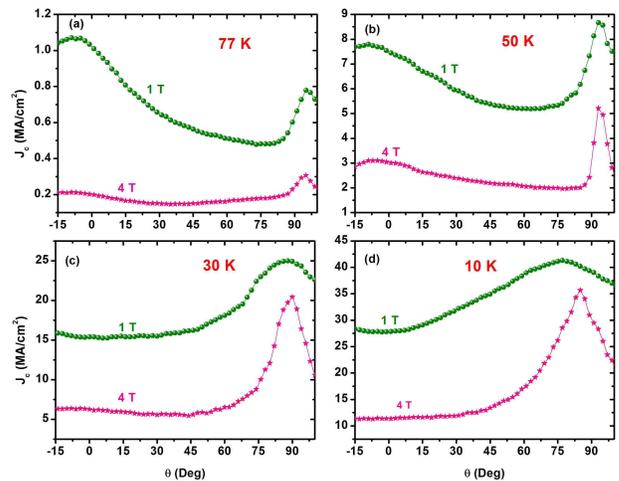}
\caption {(color online) $J_{c}$({$\theta $}) at (a) 77 K, (b) 50
K, (c) 30 K and (d) 10 K and magnetic fields up to 4 T for $H$
parallel to the $c$-axis. The $c$-axis $J_{c}$ peak caused by the
BZO nanorods is evident at 77 K and 1 T but becomes progressively
less obvious with increasing magnetic field or decreasing
temperature. Below 30 K, the peak is completely unobservable at
any magnetic field.}\label{fig:fig3}
\end{figure}

Ultra high-field magnet applications at 4.2 K are an important
 area for applications of YBCO conductors, for which we have
performed studies of $J_{c}$({$\theta $}) up to 31 T.
\cite{Xu10,Braccini11} Figure 4 shows the angular dependence of
$J_{c}$ of this MOCVD sample at 4.2 K up to 25 T. Evidently, the
$c$-axis $J_{c}$ peak is totally washed out at 4 K at all fields
from 3 T up to 25 T. As noted also in the contest of data taken at
10 K, the GL-like $J_{c}(\theta)$ dependence evolves towards a
cusp-like for magnetic fields above $\sim $5 T.

\begin{figure}[htb]
\includegraphics[width=8.1cm]{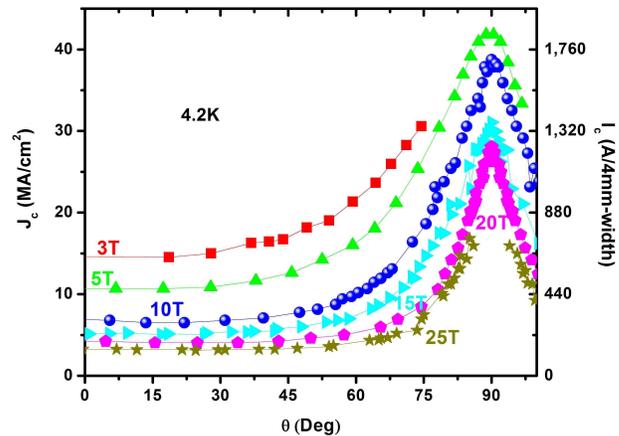}
\caption {(color online) $J_{c}$({$\theta $}) at 4.2 K under
magnetic fields up to 25 T. The $c$-axis $J_{c}$ peak is totally
absent, while the $J_{c}$ peak around the {$ab$}-plane becomes
narrower with increasing magnetic field. More importantly, this
MOCVD BZO-containing sample shows higher $J_{c}$ and a broader
$J_{c}$ around th {$ab$}-peak than MOCVD samples without BZO
nanorods.}\label{fig:fig4}
\end{figure}

\section{DISCUSSION}

The temperature and magnetic field dependence of the flux pinning
force for many superconductors often obeys a scaling relation, as
was first proposed by Fietz and Webb \cite{Fietz69, Meingast89}:
\begin{equation}
F_p (T,H)={\mbox{const}}\times[{H_{c2}(T)}]^n g(h )
\end{equation}
 in which the exponent $n$ describes the temperature dependence of the upper critical field
$H_{c2}(T)$ or in this case the irreversibility field $H_{irr}(T)$
and $g(h)$ is a pinning function which, in simple cases where the
dominant pinning mechanisms are temperature-independent, depends
only on the reduced field $h$ ($h=H$/$H_{irr}(T))$. We first
address the extent to which this simple pinning assumption is
valid for this strong pinning sample over such a broad range of
$T$ and $H$

The reduced pinning force density, $F_{p}$/$F_{p\thinspace max}$ as a
function of reduced field $h$ is plotted in Fig. 5 (a) using a
reduced field defined as $h=H$/$H_{irr}(T)$, instead of
$h=H$/$H_{c2}(T)$ for low temperature superconductors.
\cite{Meingast89} In both cases the result is the same: the
scaling field is defined for the field at which $J_{c}$ tends to
zero. At $T$ = 77 K, $F_{p}$/$F_{p\thinspace max}$ peaks at $h$ =
0.2 and $H$ = 2.0 T, which is close to the matching field of 2.6 T
for BZO nanorods in this sample. On decreasing temperature from 77
K to 50 K, the peak $h_{max}$ shifts to lower reduced field, from
0.2 to 0.15, indicating that there is no simple correlation
between $h_{max}$ and $B_{\phi }$. With further temperature
decrease to 30 K, there is no obvious change of the peak position.
However, the reduced peak width broadens as the temperature is
lowered to 40 K and then to 30 K. Thus, exact temperature scaling
does not occur in this sample, as is further demonstrated in Fig.
5 (b) by plotting log $F_{p}$ vs. log $H_{irr}(T)$ at varying
reduced fields. The scaling exponent $n$ increases from 1.65 to
2.06 as the reduced field rises from 0.25 to 0.75. However,
excellent temperature scaling is observed at constant reduced
field, as shown by the linear dependence of log $F_{p}$ on log
$H_{irr}(T)$. Taken together, both figures indicate that the
dominant pinning mechanism(s) are varying with temperature.

\begin{figure}[htb]
\includegraphics[width=8.1cm]{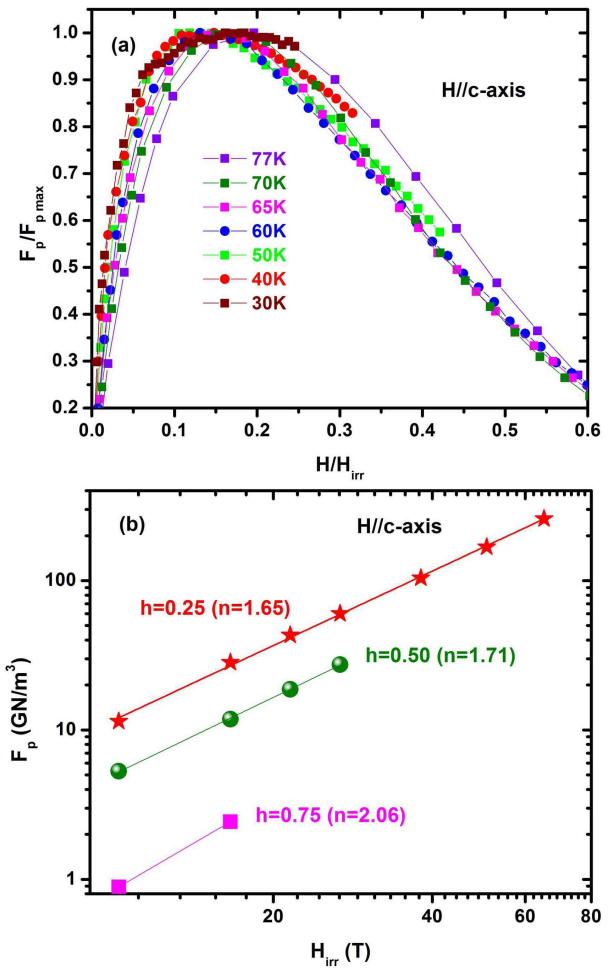}
\caption {(color online) (a) Reduced pinning force,
$F_{p}$/$F_{p\thinspace max}$ for $H$ parallel to the $c$-axis as
a function of reduced field $H$/$H_{irr}$ at temperature from 30 K
to 77 K showing the lack of a perfect temperature scaling of
$F_{p}$. (b) The scaling of $F_{p}$ as a function of
irreversibility field at several reduced magnet fields shows good
temperature scaling at constant reduced field, although the
scaling exponent is not independent of reduced
field.}\label{fig:fig5}
\end{figure}

More compelling evidence for the temperature dependence of the
pinning mechanisms is provided by the temperature dependence of
$J_{c}$/$J_{d}$, the ratio of the measured pinning critical
current density to the calculated depairing current density, which
may be used as a convenient measure of the effectiveness of the
pinning defects in the phenomenology of type II superconductors.
\cite{Blatter94} In this work, we use $J_{c}/J_{d}$ to track the
pinning strength variation with temperature. Figure 6 presents
$J_{c}/J_{d}$ for $H$ parallel to the $c$-axis as a function of
temperature at several different magnetic fields, where $J_{d}$ is
obtained from the following equation \cite{Boyd66,Lang07}

\begin{equation}
J_d(T,H)=J_d(0,
0)\left(1-\frac{T}{T_c}\right)^{3/2}\left(1-\frac{H}{H_{c2}(T)}\right)^{3/2}
\end{equation}
Here $J_{d}(0, 0) = 300$~MA/cm$^{2}$ \nocite{Lang07} (Ref.~38) and
$H_{c2}(T)$ = $H_{c2}(0)[1-(T/T_{c})^{2}]$. $T_{c} = 91$~K and
$H_{c2}(0,H||c) = 121$~T. \cite{Sekitani04} In the more often used
self-field limit, this expression reduces to the usual equation
$J_{d} = \phi_{0}/3\sqrt{3}\pi \mu_{0}\lambda^{2}\xi $.

Two evident features are observed in Fig. 6. First, it is clear
that the ratio $J_{c}$/$J_{d}$ continuously declines with
increasing temperature for all fields evaluated from $1-16$~T,
signaling that thermal fluctuations are important even in this
strong pinning regime.
\cite{Vargas01,Gurevich07,Feigel'man90,Senoussi88} However, the
dominant regime for thermal fluctuations appears to vary with
magnetic field. At the lowest field evaluated, 1 T, it appears
that three temperature regimes can be observed. On raising the
temperature from 4.2 K, there is an initially steep drop of
$J_{c}/J_{d}$ up to about 30 K, the rate of decrease then
flattening between about 30 and 65 K, before finally falling off
more rapidly again at the highest temperatures. The transition to
strong thermal fluctuations regime seems to occur at $\sim $65 K
at 1 T, 60 K at 4 T, 50 K at 8 T and 40 K at 16 T. The steep, low
temperature fall in $J_{c}$/$J_{d}$ appears at all fields, which
suggests that an additional pinning mechanism operating only at
low temperatures is present. However, the contribution of this
pinning mechanism is strongly suppressed  by both increasing
temperature and increasing field, so that at 16 T its effect is
only weakly visible as a point of inflection at $10-15$~K on the
$J_{c}$/$J_{d}(T)$ plot.

\begin{figure}[htb]
\includegraphics[width=8.1cm]{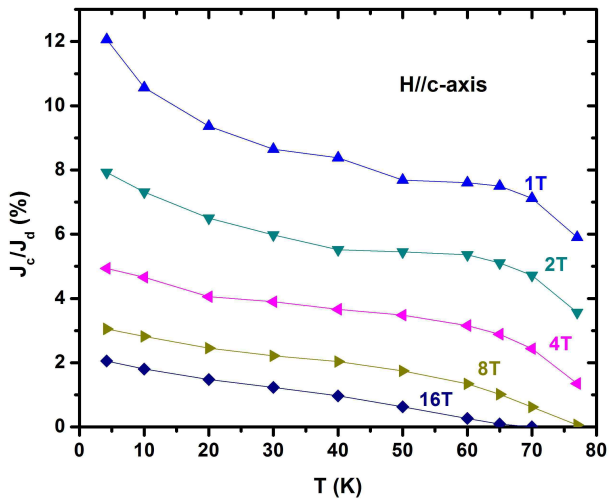}
\caption {(color online) The ratio of transport and depairing
current densities $J_{c}$/$J_{d}$ for $H$ parallel to the $c$-axis
as a function of temperature at fields of 1-16 T. Three
temperature regimes of $J_{c}(H, T)$ are visible. starting from
low temperature, $J_{c}(H, T)$ falls steeply before moderating
between about 30 to 60 K (at 1 T), then falling off again very
steeply at $\sim 60-80$~K.}\label{fig:fig6}
\end{figure}

More details of the low temperature pinning and high temperature
pinning can be obtained by plotting log $J_{c}$ vs. $T$ and log
$J_{c}$ vs. $T^{2}$. Flux pinning can be categorized as strong or
weak based on the extent of the distortion of the flux line
lattice by the pinning defects. It has been shown that the $J_{c}$
determined by weak pinning mechanisms decays as an exponential
function of temperature while $J_{c}$ of strong pinning decays as
an exponential function of $T^{2}$.
\cite{Senoussi88,Christen93,plain02,Puig08,polat11} Accordingly,
Fig. 7 (a) plots log $J_{c}$ vs. $T^{2}$. $J_{c}$ indeed decays as
an exponential function of $T^{2}$ in the intermediate temperature
regime of $30-65$~K at 1 T, $30-60$~K at 4 T, $30-50$~K at 8 T and
12 T, but it is also clear that there is an "excess" $J_{c}$ at
the lowest temperatures which quickly disappears at high
temperature. Consistent with plot of $J_{c}$/$J_{d}$ vs. $T$ in
Fig. 6, we also observe a sharp drop of $J_{c}$ at high
temperature. At lower temperatures, as shown in Fig. 7 (b), the
exponential decay of $J_{c}$, shown by the linear dependence of
log $J_{c}$ on $T$, is observed below $\sim $40 K at 1 T and 4 T
and below $\sim 30$~K 8 T and 12 T. This analysis clearly suggests
that additional weak pinning is present below $\sim 30$~K. In the
$\sim 30-60$~K range, stronger pinning centers then control
$J_{c}$ until even these strong pins are rendered ineffective by
increasing thermal fluctuations at higher temperatures. The strong
pinning provided by the BZO nanorods and RE$_{2}$O$_{3}$
precipitates in this sample significantly raise the crossover
temperature at which thermal fluctuations dominate from $\sim $60
K, as compared to $\sim $30 K in YBCO single crystal samples.
\cite{Feigel'man90} The BZO nanorods and RE$_{2}$O$_{3}$
precipitates provide strong pinning at $\sim 30-60$~K as clearly
shown by multiple other studies.
\cite{Gutierrez07,Puig08,Maiorov09}

\begin{figure}[htb]
\includegraphics[width=8.1cm]{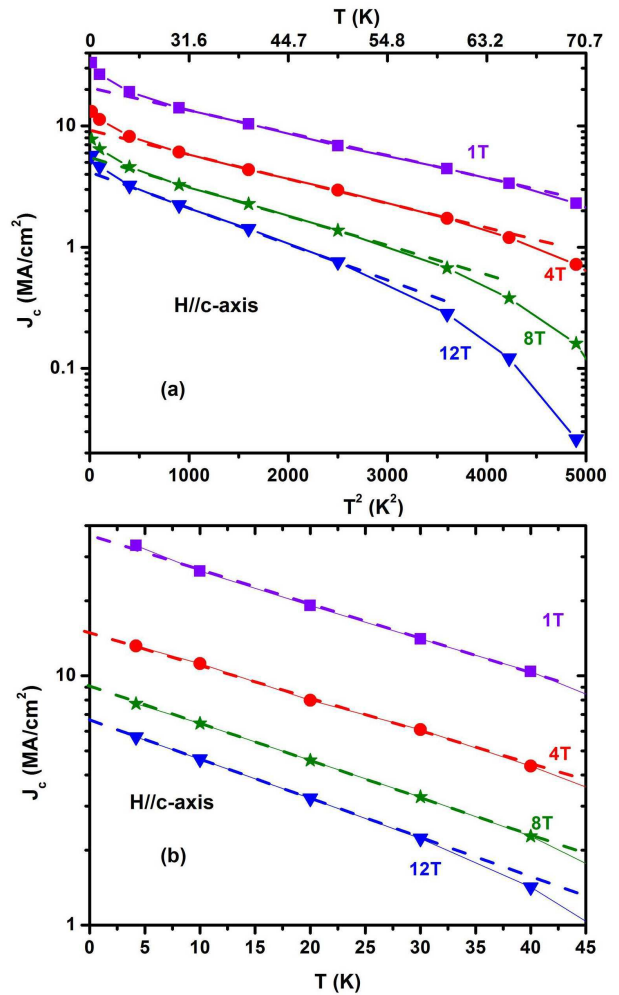}
\caption {(color online) Plot of (a) log $J_{c}$ vs. $T^{2}$ and
(b) log $J_{c}$ vs. $T$ for $H$ parallel to the $c$-axis. A linear
dependence of log $J_{c}$ on $T^{2}$ characteristic of strong
pinning is observed in the intermediate temperature range $\sim
30-60$~K in (a) while the linear dependence of log $J_{c}$ on $T$
characteristic of weak pinning is observed at low temperature
below $\sim $30 K in (b). Taken together the plots indicate that
weak pinning is adding to the strong pinning provided by
RE$_{2}$O$_{3}$ and BZO up to about 30 K and that the strong pins
then dominate $J_{c}$ until strong thermal fluctuations take over
above about 65~K (at 1T).}\label {fig:fig7}
\end{figure}

Further information about the additional low temperature weak
pinning can be obtained by analysis of the angular dependence of
$J_{c}$ at 4.2 K using the anisotropic scaling approach
\cite{Blatter94} proposed by Civale {et al.} \cite{Civale04} If
the pinning is due to uncorrelated defects randomly distributed
over angular space, then $J_{c}$ should depend on $H$ and {$\theta
$} only through a single variable
$J_{c}(H,\theta)=J_{c}[H\varepsilon(\theta)]$ where $\varepsilon
(\theta)=[{\cos}^{2}(\theta)+\gamma^{-2}{\sin}^{2}{(\theta)}]^{1/2}$,
where $\gamma^{2}$ is the electronic mass anisotropy parameter.
Since we have observed that $J_c$  follows $J_c\propto H^{-\alpha
}$ at low temperature especially at 4.2 K, as shown in Fig. 2 (a),
then $J_{c}(H,\theta)\propto [H\varepsilon(\theta)]^{-\alpha}$.
Power law behavior with $\alpha \approx  0.7$ for BZO-containing
samples and $\alpha  \approx  0.5$ for non-BZO samples was
reported by us previously. \cite{Braccini11} Figure 8 presents
$J_{c}(\theta )$ at 5, 15 and 25~T using the anisotropic scaling
parameters $\alpha= 0.7$ and $\gamma=3$. These curves give a
reasonable fit of the data in the low angle region, from
$0^{\circ}$ up to about $60-70^{\circ}$, leading us to conclude
that, over this wide range all around the $c$-axis where the
correlated effects of the BZO nanoparticles are so evident at
higher temperatures, the pinning is indeed uncorrelated and
random. Clearly the most reasonable interpretation of this
behavior is that the {\em additional} pinning that operates only
up to $30-40$~K dominates over the strong pinning effect of the
BZO nanorods at lower temperatures. It is interesting that the
scaling suggests an effective anisotropy $\gamma=3$ that is less
than the $H_{c2}$ anisotropy = 5 expected from the intrinsic mass
anisotropy $\gamma^{2}=25-30$. The reasons for this behavior are
unclear at this time, but smaller anisotropy is positive in any
case of applications. It is interesting that $\gamma=3$ is also an
excellent fit to the very broad study of the angular variation of
$H_{irr}(T)$ for a very similar sample over the range about 55-80
K and $H$ up to 45 T. \cite{Tarantini11} In this case the
effective $\gamma$ was evaluated in the limit that the pin
strength goes to zero, rather than in the very finite pin strength
limit assessed here, implying a rather consistent behavior over a
wide range of the superconducting phase space.

The power law dependence of $J_{c}$ on $H$ has been reported in
many works \cite{ Maiorov09,Zuev08}. However, the increase of
{$\alpha $} from 0.5 (no nanorods) to 0.7 at 4.2 K when BZO
nanorods are present is in contrast to high temperature
observations where the presence of BZO nanorods decreases {$\alpha
$}. \cite{Braccini11,Maiorov09} This is yet more evidence that
different pinning mechanisms operate at high and low temperatures.

Based on the above analysis, we propose that BZO additions lead to
a significant density of strain-induced, weak pins that cannot
resist thermal fluctuation much above about 30 K. Since they do
operate at all fields up to 31 T at 4.2 K, it is clear that they
are much denser than the strong BZO and RE$_{2}$O$_{3}$ pins. Due
to the strong effects of increasing field and temperature, we
assume that these highly effective pins are dense but point pins
(effective size of the order of $\xi^3$ or less, where $\xi$ is
the coherence length). Our earlier 4.2 K comparison of non-BZO and
BZO-containing samples up to 31 T showed two significant features:
One was that the BZO samples have significantly higher $J_{c}$
while the second was that the enhancement disappeared by
$30-35$~T, certainly a high field but actually only about 25~{\%}
of $H_{irr}$ or $H_{c2}$ at 4.2 K. Here our variable temperature
examination of the BZO sample shows that the uncorrelated pinning
effects produced by the point pins are only visible up to $\sim
30$~K. Both characterizations show that the pin strength decays
rapidly with increasing $H$ and $T$, consistent with them being
small and easily thermally depinned. As Fig. 1 shows, only the
larger BZO and RE$_{2}$O$_{3}$ strong pins are visible in TEM, so
we do not yet have a measure of the point pin concentration. But
we can reasonably infer that their density must be high, since
they are able to completely hide the effect of the BZO nanorods,
even at fields below $B_{\phi }$ (compare the $J_{c}$(1 T) data at
77 and 10 K in Figs. 3a and 3d). Indeed, the depressed $T_{c}$ of
MOCVD and PLD YBCO samples induced by BZO nanorods has recently
been attributed to oxygen deficiencies introduced by strain
imposed by the lattice mismatch between BZO nanorods and the YBCO
matrix. \cite{Cantoni11} The specific supporting evidence was
provided by atomic-resolution Z-contrast imaging and electron
energy loss spectroscopy which showed oxygen deficiencies
surrounding BZO nanorods, a finding which also supports our
proposal for the presence of dense point pins. Thus the overall
conclusion of our study is extremely positive: the splayed BZO
nanorods found in this coated conductor provide strong correlated
pinning to enhance $J_{c}$ around $c$-axis, a result shown in
numerous studies in the 50-77 K range, but they also greatly add
to the lower temperature $J_{c}$ by additionally inducing dense
but weak isotropic pinning by strain-induced point defects that
raise $J_{c}$ in the whole angular range at fields up to at least
31 T at 4.2 K where thermal depinning effects are small.

\section{CONCLUSIONS}

In this paper, we presented a very detailed $J_{c}(H, T, \theta)$
characterization of a modern, very high critical current density
REBCO thin film containing $\sim c$-axis oriented BZO nanorods and
{$ab$}-plane RE$_{2}$O$_{3}$ pinning arrays over an exceptionally
broad temperature ($4.2-77$ K) and magnetic field range ($0-31$
T). Analyzing the $J_{c}$ data we studied the pinning evolution on
temperature associated with BZO nanorods An important new
conclusion is that weak isotropic pinning from point defects
produced by the strain field around BZO nanorods dominates $J_{c}$
at low temperatures More specifically, we observed that the usual
 $c$-axis $J_{c}$ peak caused by BZO nanorods disppears with
decreasing temperature, and vanishes completely below $\sim 30$~K.
At 4.2 K, we found that $J_{c}$ along the $c$-axis decays as
$J_{c}  \propto H^{-\alpha}$ with magnetic field up to 31 T.
Although $J_{c}$ decays faster with magnetic field compared with
samples without BZO nanorods at 4.2 K, it is still higher at field
up to at least 31 T. At low tempertures, the $c$-axis $J_{c}$ peak
is not seen at any magnetic field and the only correlated pinning
present occurs around the {$ab$}-plane.

\begin{figure}[htb]
\includegraphics[width=8.1cm]{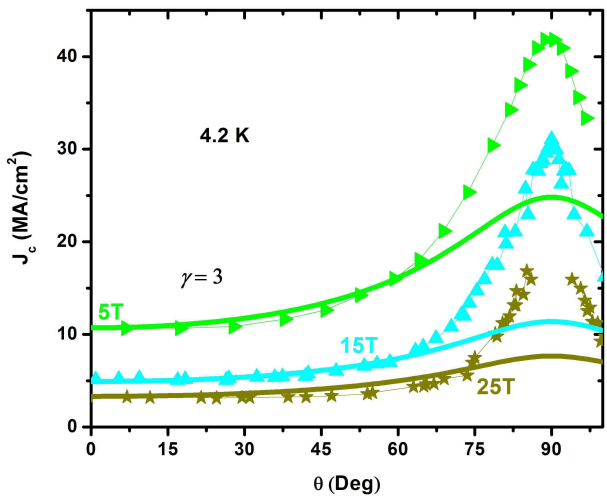}
\caption {(color online) Angular dependence of $J_{c}$ (symbols)
and $J_{c}$ calculated from anisotropic G-L scaling model (thick
lines) at 4.2 K and magnetic fields 5, 15 and 25 T. Anisotropic
G-L scaling describes well the experimental data over a broad
angular range up to, $\sim  0-60^{\circ}$ with $\alpha  = 0.7$ and
$\gamma  = 3$. Thus, the weak uncorrelated pinning contributes to
$J_{c}$ at 4.2 K except in the vicinity of the {$ab$}-plane where
intrinsic pinning is important.}\label{fig:fig8}
\end{figure}

\acknowledgments

We are very grateful to colleagues in the HTS R{\&}D group at the
NHMFL who have provided many valuable comments and discussions,
especially Alex Gurevich, David Hilton, Fumitake Kametani, and
Chiara Tarantini. Some aspects of this work were supported by the
Department of Energy, Office of Electric Delivery and Energy
Research (grant number: DE-FC07-08ID14916) and some by the
National High Magnetic Field Laboratory, which is supported by NSF
Cooperative Agreement DMR-0654118 and by the State of Florida.


\end{document}